\documentclass[a4paper,10pt]{article}
\usepackage{graphicx}
\usepackage{amssymb}
\usepackage{amsmath}
\usepackage{nicefrac}
\usepackage{dcolumn}
\usepackage{multirow}
\usepackage{cite}
\usepackage{mathrsfs}
\usepackage{bm}
\usepackage[usenames,dvipsnames]{xcolor}
\usepackage[colorlinks=true,citecolor=Blue,linkcolor=RubineRed,urlcolor=Blue]{hyperref}
\usepackage{bbm}
\usepackage{hyperref}
\usepackage{comment}
\usepackage{amsfonts}
\usepackage{float}
\usepackage{caption}
\usepackage{subcaption}

\usepackage{xcolor}
\usepackage{soul}
\setstcolor{red}

\begin{document}

\def\anglefig{0}
\def\scalefig{0.90}

\huge

\begin{center}
On the Thomas-Fermi model: Gabor J. Kalman's contribution and numerical approximations
\end{center}

\vspace{0.5cm}

\large

\begin{center}
Jean-Christophe Pain$^{a,b,}$\footnote{jean-christophe.pain@cea.fr}
\end{center}

\normalsize

\begin{center}
\it $^a$CEA, DAM, DIF, F-91297 Arpajon, France\\
\it $^b$Universit\'e Paris-Saclay, CEA, Laboratoire Mati\`ere sous Conditions Extr\^emes,\\
\it 91680 Bruy\`eres-le-Ch\^atel, France
\end{center}

\vspace{0.5cm}

\abstract{In this article, we would like to pay tribute to Gabor Kalman, outlining his contribution to a model widely used in dense plasma physics: the high-temperature Thomas-Fermi model. The approach of Ruoxian Ying and Kalman relies on the separation of the bound and free electrons, a physically reasonable definition of the bound electrons, a description of the source density in the Poisson equation through the electron-ion and ion-ion pair correlation functions and a determination of the degree of ionization from the minimization of the total free energy. We also report on different approximations of the function $\Phi$, which is a cornerstone of the original Thomas-Femi model.}

\section{Introduction}\label{sec1}

The theoretical determination of the ionization state of partially ionized plasmas must take into account the effects of density on bound states and strong ionic coupling. The corresponding methods boil down to two main categories. Chemical-picture methods consider that the system consists of distinct chemical species; it is thus necessary to account for the effect of the plasma environment on their internal states. On the other hand, physical-picture methods describe the plasma in terms of its fundamental constituents, i.e., electrons and nuclei, so that plasma effects on bound states are a very important feature of the modeling \cite{Rogers2000}. One of the main issues of the Saha-equation approach is the self-consistent determination of the bound-state energies in the presence of the many-body plasma environment \cite{Ebeling1988}. Most of the work done along these lines has been mainly restricted to Debye screening. Besides the static screening, the Debye dynamic screening has been treated as well \cite{Ropke1978,Zimmerman1978}, e.g., for solving the Bethe-Salpeter equation \cite{Bethe1951}, describing the bound states of a two-body quantum field theoretical system in a relativistically covariant formalism. The problem of boud/free states also occurs in warm dense matter, where a proper distinction is rather involved. New approaches based on density functional theory are given in Ref. \cite{Bethkenhagen2020}. The average-atom approach — in addition to the Thomas-Fermi calculations discussed here — has been explored more thoroughly (see for instance Ref. \cite{Wetta2022}). The Thomas-Fermi statistical model is relatively simple and has proven to be a powerful method to calculate average properties of the plasma, such as its equation of state and mean ionization. The model, originally formulated for isolated neutral atoms, has been extended in several ways to describe hot and dense plasmas \cite{Feynman1949,Latter1955,Latter1956}. A rather exhaustive review of Thomas-Fermi approach and its variations is provided in Ref. \cite{Spruch1991}. Kobayashi solved the zero-temperature Thomas-Fermi equation for positive ions still in the framework of the isolated-ion model, and calculated the degree of ionization for arbitrary atomic number as a function of the ionic radius \cite{Kobayashi1959}. Ivanov studied the asymptotic solution of Thomas-Fermi equation \cite{Ivanov1958}. Feng used a screened potential given by the Debye-H\"uckel model \cite{Feng1981}. Dharma-Wardana and Perrot applied the density-functional theory to hydrogenic plasmas \cite{Dharma1982,Dharma1987} and achieved unambiguous treatment of the somewhat intuitive corrections to the Thomas-Fermi equation arising from the ion distribution and proposed by More and Skupsky \cite{More1976}. Rozsnyai generalized the Thomas-Fermi approach to mixtures (multicomponent plasmas) \cite{Rozsnyai1976,Rozsnyai1977} and de Carvalho to the relativistic case \cite{deCarvalho2014}.

Ruoxian Ying and Kalman established a new model based on the Thomas-Fermi method \cite{Thomas1927,Fermi1927,Torrens1972} in order to calculate the average degree of ionization of a plasma \cite{Kalman1983,Ying1986,Ying1989}. This new model is characterized by the following features: 

\begin{itemize}
    \item (i) the bound electrons and free electrons are treated separately, 
    \item (ii) a physically reasonable definition of the bound electrons is chosen; the bound-electron density is given by a momentum cutoff integral of the Fermi-Dirac distribution (this ensures that the bound-electron density vanishes at the ion boundary and that the maximum energy of the bound electrons is the same everywhere within the ion),
    \item (iii) the system is described as a strongly coupled plasma of free electrons and Thomas-Fermi ions, 
    \item (iv) the density in the Poisson equation is determined by the electron-ion and ion-ion correlation functions, and 
    \item (v) the degree of ionization is calculated through the minimization of the total free energy. 
\end{itemize}

Ruoxian Ying and Kalman investigated different descriptions of the ionic correlations. In this paper, we consider the simplest version of their work, representing the ion-ion correlation function by a Heaviside distribution. The main features of their model are recalled in section \ref{sec2}. The concept of chemical potential of bound electrons, together with comparisons with the results from the Saha equation, are presented in section \ref{sec3}. Although this goes beyond the scope of Kalman's work, in section \ref{sec4} we discuss different approximations of the function $\Phi$, which is key ingredient of the Thomas-Fermi original model. The different ideas underlying such approximations, however, could be generalized to the $\Psi$ function entering the model of Ruoxian Ying and Kalman.

\section{Main outlines of the formalism}\label{sec2}

Let us denote $z$ the mean ionization (average number of free electrons per atom, determined in a self-consistent way), $n_b$, $n_f$ and $n_i$ the radial bound-, free- and ion-densities respectively. The total central-field potential felt by an electron is obtained through the Poisson equation \cite{Kalman1983,Ying1986,Ying1989}: 
\begin{equation}
    \nabla^2V(r)=4\pi e\left\{n_b[V(r)]+n_f[r,V_1(r)]-zn_i[r,V_1(r)]\right\},
\end{equation}
where $V_1$ is the potential due to the bound electrons:
\begin{equation}
    \nabla^2V_1(r)=4\pi en_b[V(r)].
\end{equation}
In the finite-temperature Thomas-Fermi model version of Ruoxian Ying and Kalman, the bound-electron density reads
\begin{equation}\label{nb}
    n_b[V(r)]=\frac{8\pi}{h^3}\int_0^{p_m}\frac{p^2}{\exp\left\{\left[\displaystyle\frac{p^2}{2m_e}-eV(r)-\alpha\right]/(k_BT)\right\}+1}dp,
\end{equation}
where $m_e$ represents the electron mass and $\alpha$ is a parameter to be determined. The upper bound of the integral is
\begin{equation}
    p_m=\left\{2m_e e[V(r)-V(r_0)]\right\}^{1/2},
\end{equation}
$r_0$ being the ion-sphere radius. The free electron density is given by
\begin{equation}
    n_f[r,V_1(r)]=\bar{n}_f\left\{1+g_{ei}[r,V_1(r)]\right\}
\end{equation}
and the bound-electron density by
\begin{equation}
    n_i[r,V_1(r)]=\bar{n}_i\left\{1+g_{ii}[r,V_1(r)]\right\},
\end{equation}
where $g_{ei}$ and $g_{ii}$ are the electron-ion and ion-ion correlation functions (radial distribution functions) respectively, and $\bar{n}_i$ and $\bar{n}_f=z\bar{n}_i$ the average densities of ions and free electrons. The authors first make the approximation $g_{ei}=g_{ii}=0$ and $n_f=\bar{n}_f$, $n_i=\bar{n}_i$ yielding
\begin{equation}
    \nabla^2V(r)=4\pi en_b[V(r)]=4\pi e\frac{8\pi}{h^3}\int_0^{p_m}\frac{p^2}{\exp\left\{\left[\displaystyle\frac{p^2}{2m_e}-eV(r)-\alpha\right]/(k_BT)\right\}+1}dp,
\end{equation}
with $V(r_0)=z^2/r_0$, $V'(r_0)=-z^2/r_0^2$ and $V(r)\approx (Ze)/r$ as $r\rightarrow 0$. The quantity $z=4\pi n(r_0)r_0^3/3$ represents the free-electron number, $n(r)$ being the total electron density. The free-electron density reads
\begin{equation}
    n_f=\frac{4\pi}{h^3}(2m_ek_BT)^{3/2}I_{1/2}\left(\frac{\alpha}{k_BT}\right),
\end{equation}
where
\begin{equation}
    I_n(x)=\int_0^{\infty}\frac{y^n}{1+e^{y-x}}dy=-n!\,\mathrm{Li}_{n+1}(-e^x),
\end{equation}
where $\mathrm{Li}_{n}(x)$ represents the polylogarithm of order $n$ \cite{Lewin1981}:
\begin{equation}
    \mathrm{Li}_{s}(z)=\sum _{k=1}^{\infty }{z^{k} \over k^{s}}=z+{z^{2} \over 2^{s}}+{z^{3} \over 3^{s}}+\cdots
\end{equation}
This definition is valid for arbitrary complex order $s$ and for all complex arguments $z$ with $|z| < 1$; it can be extended to $|z| \geq 1$ by the process of analytic continuation. One has
\begin{equation}
    \mathrm{Li}_{s+1}(z)=\int _{0}^{z}\frac{\mathrm{Li}_{s}(t)}{t}dt,
\end{equation}
and thus the dilogarithm is an integral of a function involving the logarithm, \emph{etc.} For non-positive integer orders $s$, the polylogarithm is a rational function.
\begin{equation}
\mathrm{Li}_{s}(z)=\sum _{k=0}^{\infty }(-1)^{k}(1-2^{1-2k})(2\pi )^{2k}{B_{2k} \over (2k)!}{[\ln(-z)]^{s-2k} \over \Gamma (s+1-2k)},
\end{equation}
where $B_{2k}$ are the Bernoulli numbers. $I_n$ is referred to as a Fermi integral \cite{Dingle1957}. 

The free energy of bound electrons can be put in the form
\begin{eqnarray}
    F_1&=&(Z-z)\alpha-\frac{2}{3}\frac{aZk_BT}{\psi(0)}\ln\left\{1+\exp[\psi(1)]\right\}\int_0^1dxx^2\left[\frac{\psi(x)}{x}-\psi(x)\right]^{3/2}\nonumber\\
    & &+\frac{1}{3}\frac{aZk_BT}{\psi(0)}\int_0^1dx xJ_{1/2}\left[\frac{\psi(x)}{x},\psi(1)\right]\left[\psi(x)-\psi'(1)x-2\psi(0)\right],
\end{eqnarray}
where
\begin{equation}
    a=\frac{(4\pi e)^2(2m_e)^{3/2}}{h^3}(k_BT)^{1/2}r_0^2
\end{equation}
and
\begin{equation}
\psi(x)=\frac{[\alpha+eV(r)]r}{k_BTr_0}=\frac{[\alpha+eV(xr_0)]x}{k_BT}
\end{equation}
as well as
\begin{equation}
    J_n(x,x_0)=\int_0^{x-x_0}\frac{y^n}{1+e^{y-x}}dy.
\end{equation}
The function $\Psi(x)$ satisfies the non-linear differential equation
\begin{equation}\label{tfk}
    \Psi''(x)=axJ_{1/2}\left[\frac{\Psi(x)}{x},\Psi(1)\right],
\end{equation}
with the boundary conditions
\begin{equation}
    \Psi(1)=\frac{\alpha+ze^2/r_0}{k_BT},\qquad \Psi'(1)=\frac{\alpha}{k_BT}, \qquad \mathrm{and} \qquad \Psi(0)=\frac{Ze^2}{k_BTr_0}.
\end{equation}
Note that, in the case of a pure Coulomb potential, the first integral in the right-hand side can be expressed in terms of elliptic $E$ and $K$ integrals \cite{Byrd1971}. Conversely, the free energy of free electrons reads
\begin{equation}
    F_2=zk_BT\left[\frac{\mu}{k_BT}-\frac{2}{3}\displaystyle\frac{I_{3/2}\left(\displaystyle\frac{\mu}{k_BT}\right)}{I_{1/2}\left(\displaystyle\frac{\mu}{k_BT}\right)}\right]
\end{equation}
and the total free energy is $F=F_0+F_1+F_2$, $F_0$ being the translational motion of ions. Then, $F$ can be minimized with respect to variables $\alpha$, $z$ and $r_0$.

\section{Chemical potential of bound electrons}\label{sec3}

The chemical potential of an ideal gas of free electrons is known to be negative at high temperature and becomes positive at low temperature. Ruoxian Ying and Kalman showed that the chemical potential of bound electrons has the same temperature dependence, which enables one to perform a minimization of the total free energy, requiring that the chemical potentials of the bound and free electrons be equal. The expression for the chemical potential of the bound electrons is given by
\begin{equation}
\mu_b=\left.\frac{\partial F_1}{\partial (Z-z)}\right|_{r_0}
\end{equation}
where the free energy is
\begin{equation}
    F=E-TS=(Z-z)\alpha-E_{ee}+k_BT\sum_i\ln(1-n_i),
\end{equation}
with
\begin{equation}
    E_{ee}=\frac{(4\pi e)^2}{2}\int_0^{r_0}r^2dr\int_0^{r_0}r'^2dr'\frac{n(r)n(r')}{|\vec{r}-\vec{r'}|}.
\end{equation}
Taking into account the fact that
\begin{equation}
    \frac{\partial E_{ee}}{\partial z}=(4\pi e)^2\int_0^{r_0}r^2dr\int_0^{r_0}r'^2dr'\frac{n(r)}{|\vec{r}-\vec{r'}|}\frac{\partial n(r')}{\partial z},
\end{equation}
Ruoxian Ying and Kalman obtain
\begin{equation}
    k_BT\frac{\partial}{\partial z}\sum_i\ln(1-n_i)=-4\pi\int_0^{r_0}r^2dr\int_0^{p_m}p^2\ln\left[1+\exp\left\{\left[eV(r)+\alpha-\frac{p^2}{2m_e}\right]/(k_BT)\right\}\right]dp,
\end{equation}
and
\begin{eqnarray}
    k_BT\frac{\partial}{\partial z}\sum_i\ln(1-n_i)&=&-4\pi\int_0^{r_0}drr^2n(r)\left[\frac{\partial eV(r)}{\partial z}+\frac{\partial\alpha}{\partial z}\right]\nonumber\\
    & &-\frac{32\pi^2k_BT}{h^3}\ln\left[1+\exp\left(\frac{\alpha+eV_0}{k_BT}\right)\right]\frac{1}{3}(2m_ee)^{3/2}\frac{\partial}{\partial z}\int_0^{r_0}drr^2\left[V(r)-V_0\right]^{3/2},
\end{eqnarray}
where we have used Eq. (\ref{nb}) for $n(r)$. Using $V(r)=V_e(r)+V_n(r)$ together with
\begin{equation}
    V_e(r)=-4\pi e\int_0^{r_0}r'^2dr'\frac{n(r')}{|\vec{r}-\vec{r}'|}
\end{equation}
and $V_n(r)=Ze/r$, one obtains
\begin{eqnarray}\label{c8}
    k_BT\frac{\partial}{\partial z}\sum_i\ln(1-n_i)&=&(4\pi e)^2\int_0^{r_0}drr^2\int_0^{r_0}dr'r'^2\frac{n(r')}{|\vec{r}-\vec{r}'|}\frac{\partial n(r')}{\partial z}-(Z-z)\frac{\partial\alpha}{\partial z}\nonumber\\
    & &-\frac{32\pi^2k_BT}{h^3}\ln\left[1+\exp\left(\frac{\alpha+eV_0}{k_BT}\right)\right]\frac{1}{3}(2m_ee)^{3/2}\frac{\partial}{\partial z}\int_0^{r_0}drr^2\left[V(r)-V_0\right]^{3/2}.
\end{eqnarray}
The chemical potential of bound electrons reads finally
\begin{eqnarray}
    \mu_b=\frac{\partial F_1}{\partial (Z-z)}&=&\alpha+\frac{32\pi^2k_BT}{3h^3}\ln\left[1+\exp\left(\frac{\alpha+eV_0}{k_BT}\right)\right]\nonumber\\
    & &\times(2m_ee)^{3/2}\frac{\partial}{\partial z}\int_0^{r_0}dr r^2\left[V(r)-V_0\right]^{3/2}.
\end{eqnarray}
At the zero-temperature limit, since $\alpha > -ze/r_0$, the total number of bound electrons is

\begin{equation}\label{vintuit}
    Z-z=\frac{32\pi^2}{3h^3}\int_0^{r_0}drr^2\left\{2m_ee\left[V(r)-V_0\right]^{3/2}\right\}
\end{equation}
and then the chemical potential reduces to
\begin{equation}
    \mu_b=\alpha-k_BT\ln\left[1+\exp\left(\frac{\alpha+eV_0}{k_BT}\right)\right]=-\frac{ze^2}{r_0},
\end{equation}
which is negative at zero temperature. In that case $-ze /r_0$ represents the Fermi energy for bound electrons. The chemical potential of free electrons, on the other hand, is positive at zero temperature and thus cannot be equal to $\mu_b$. It is worth noting that at finite temperature $\mu_b$ can be either positive or negative, depending on the value of $\alpha$. This explains why, as temperature is lowered from a high value, $\mu_b$ changes from negative to positive, and then becomes negative again at zero-temperature. 

\begin{figure}[!ht]
 \begin{minipage}[c]{0.48\textwidth}
  \centering
  \includegraphics[width=\textwidth,angle=\anglefig,scale=\scalefig]{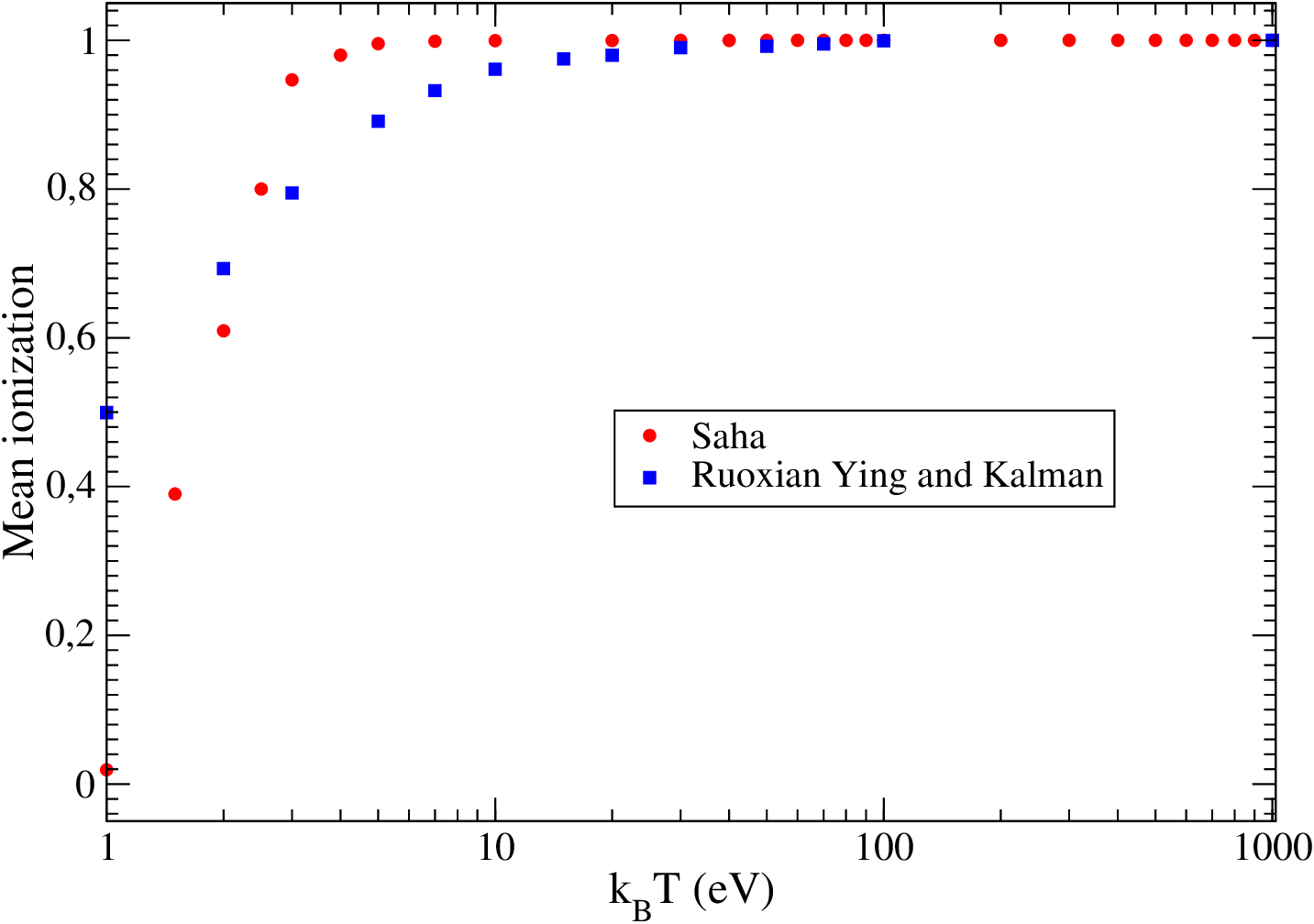}
  \caption{(Color online) Mean ionization as a function of temperature (more precisely thermal kinetic energy $k_BT$) for $n_i$=10$^{19}$ cm$^{-3}$ in the case of a hydrogen plasma.\label{kalman_1019}}
 \end{minipage}\hfill
 \begin{minipage}[c]{0.48\textwidth}
  \centering
  \includegraphics[width=\textwidth,angle=\anglefig,scale=\scalefig]{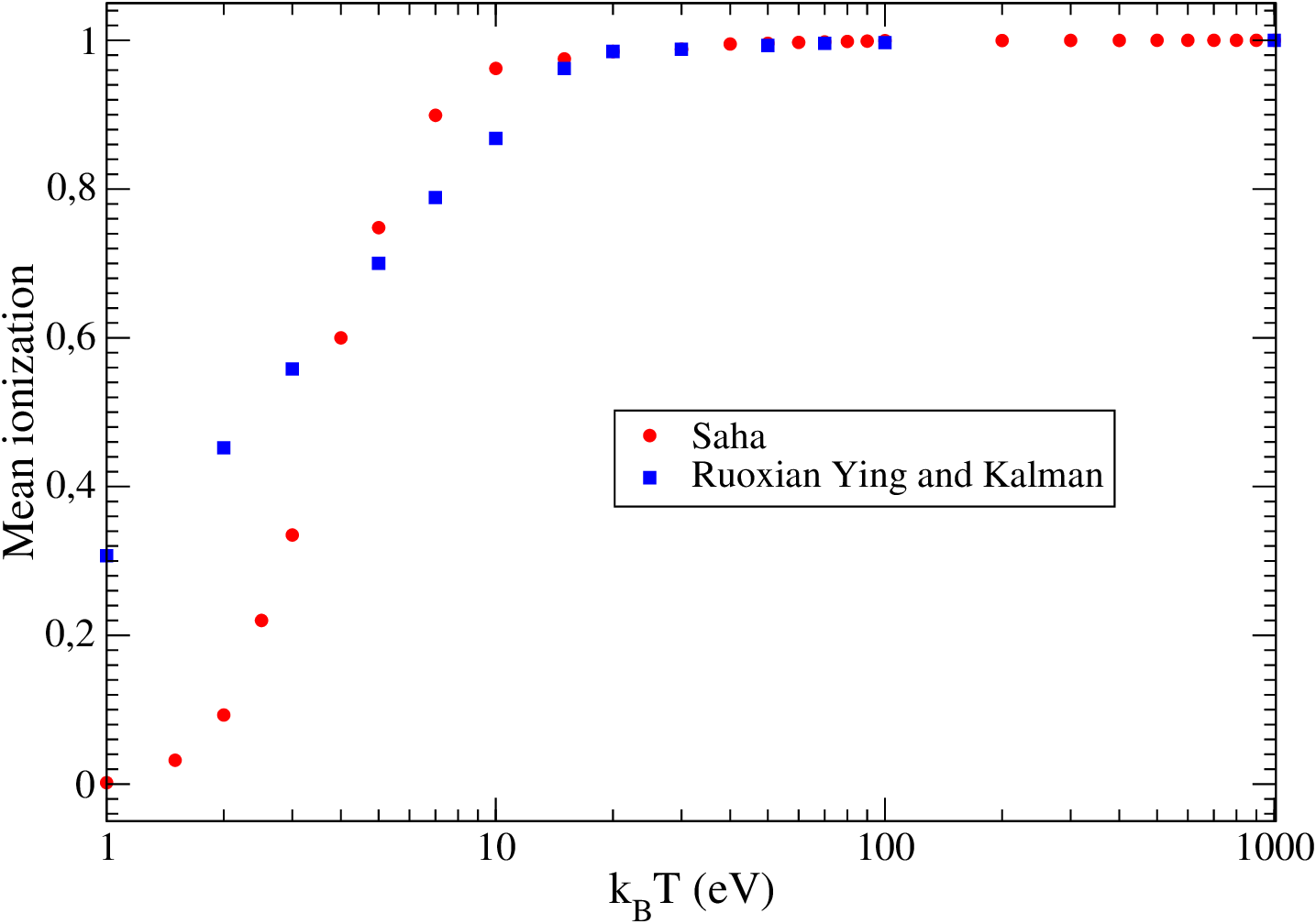}
  \caption{(Color online) Mean ionization as a function of temperature for $n_i$=10$^{21}$ cm$^{-3}$ in the case of a hydrogen plasma.\label{kalman_1021}}
 \end{minipage}
\end{figure}

\begin{figure}[!ht]
	\centering
	\includegraphics[width=8cm]{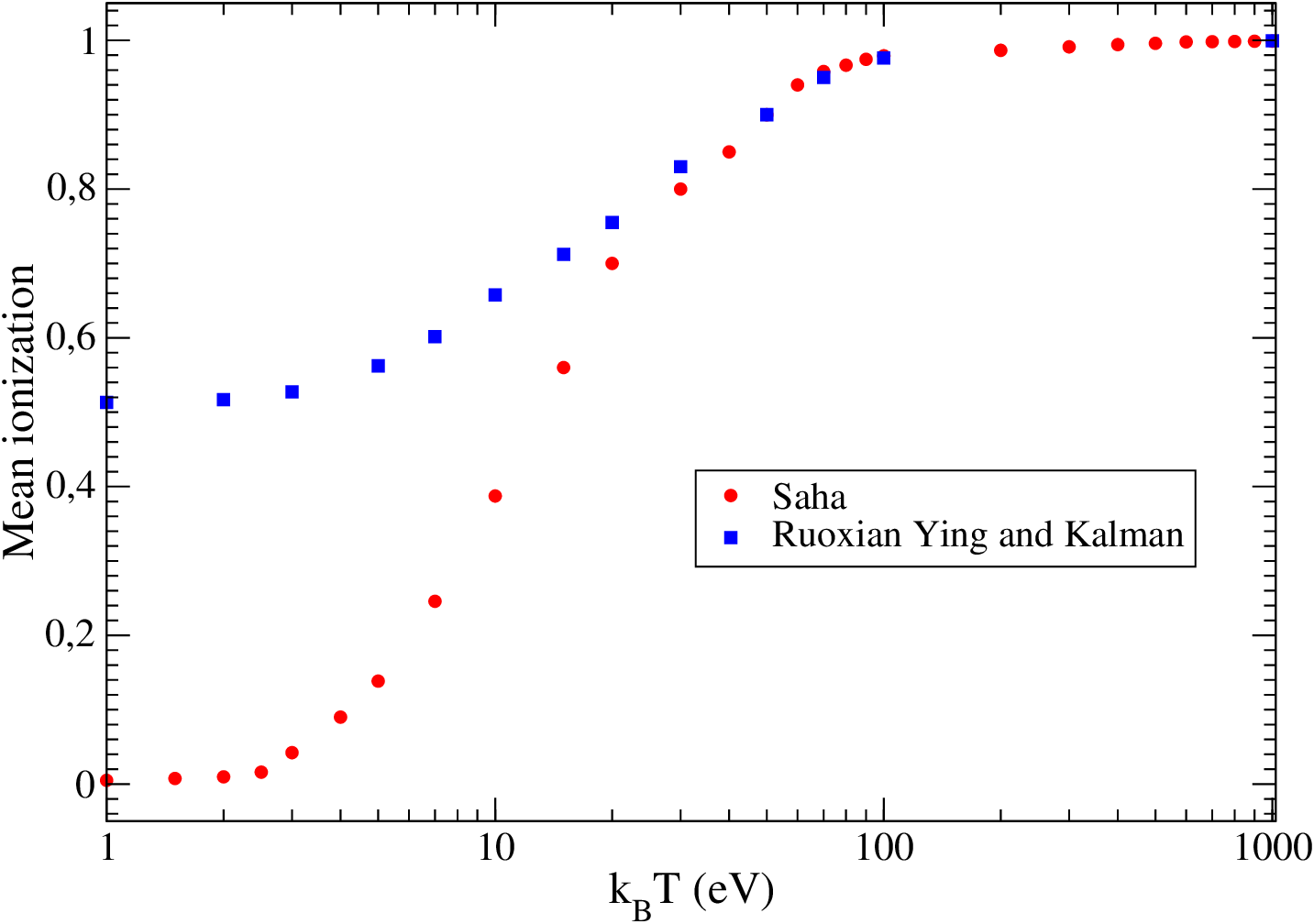} 
	\caption{Mean ionization as a function of temperature for $n_i$=10$^{23}$ cm$^{-3}$ in the case of a hydrogen plasma.}
	\label{kalman_1023}
\end{figure} 

\begin{figure}[!ht]
	\centering
	\includegraphics[width=8cm]{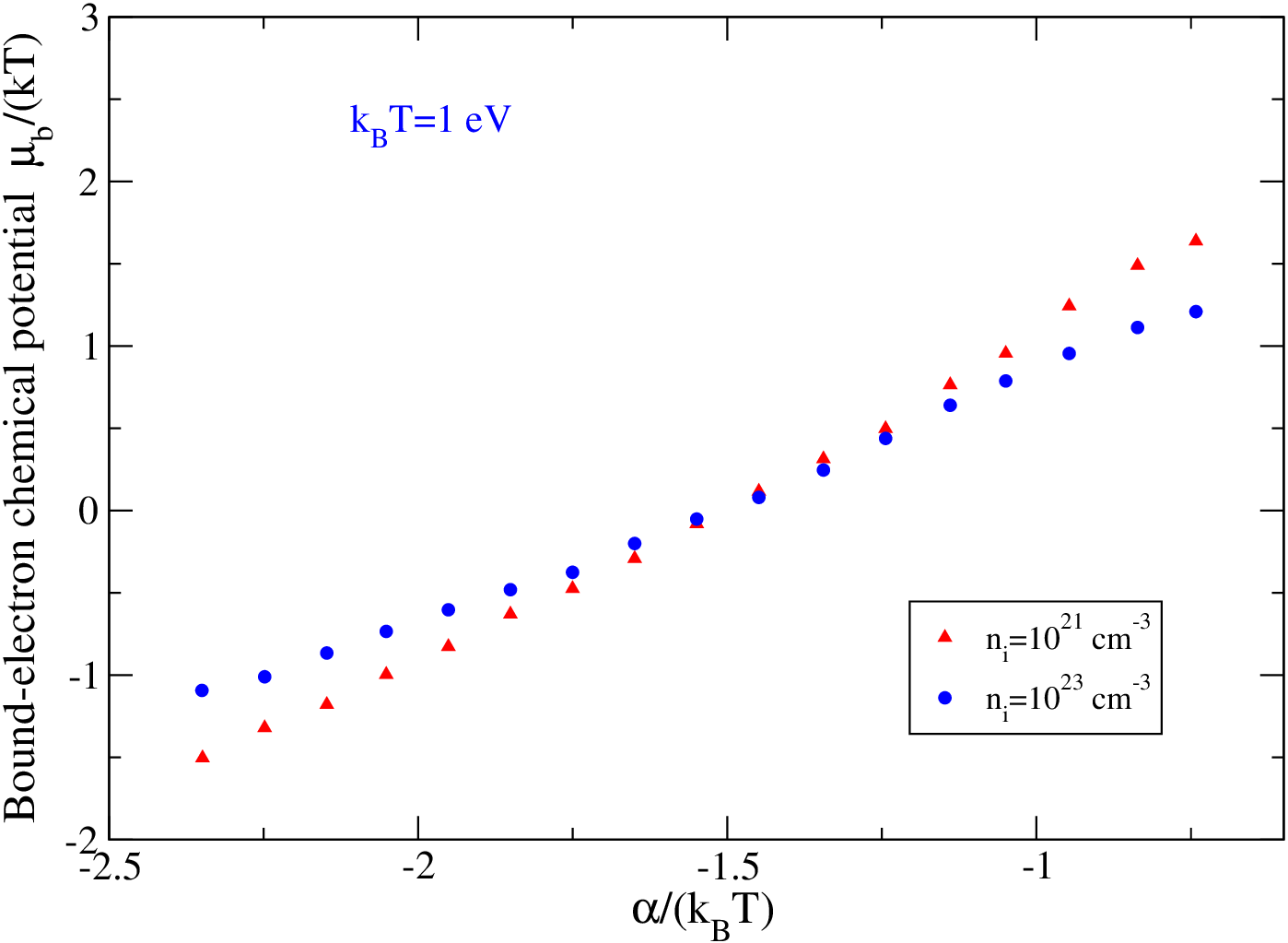} 
	\caption{Reduced bound-electron chemical potential $\mu_b/(k_BT)$ of a hydrogen plasma as a function of reduced parameter $\alpha/(k_BT)$ for $n_i$=10$^{21}$ cm$^{-3}$ and $n_i$=10$^{23}$ cm$^{-3}$. The thermal kinetic energy is $k_BT$=1 eV.}
	\label{fig17}
\end{figure} 

\begin{figure}[!ht]
	\centering
	\includegraphics[width=8cm]{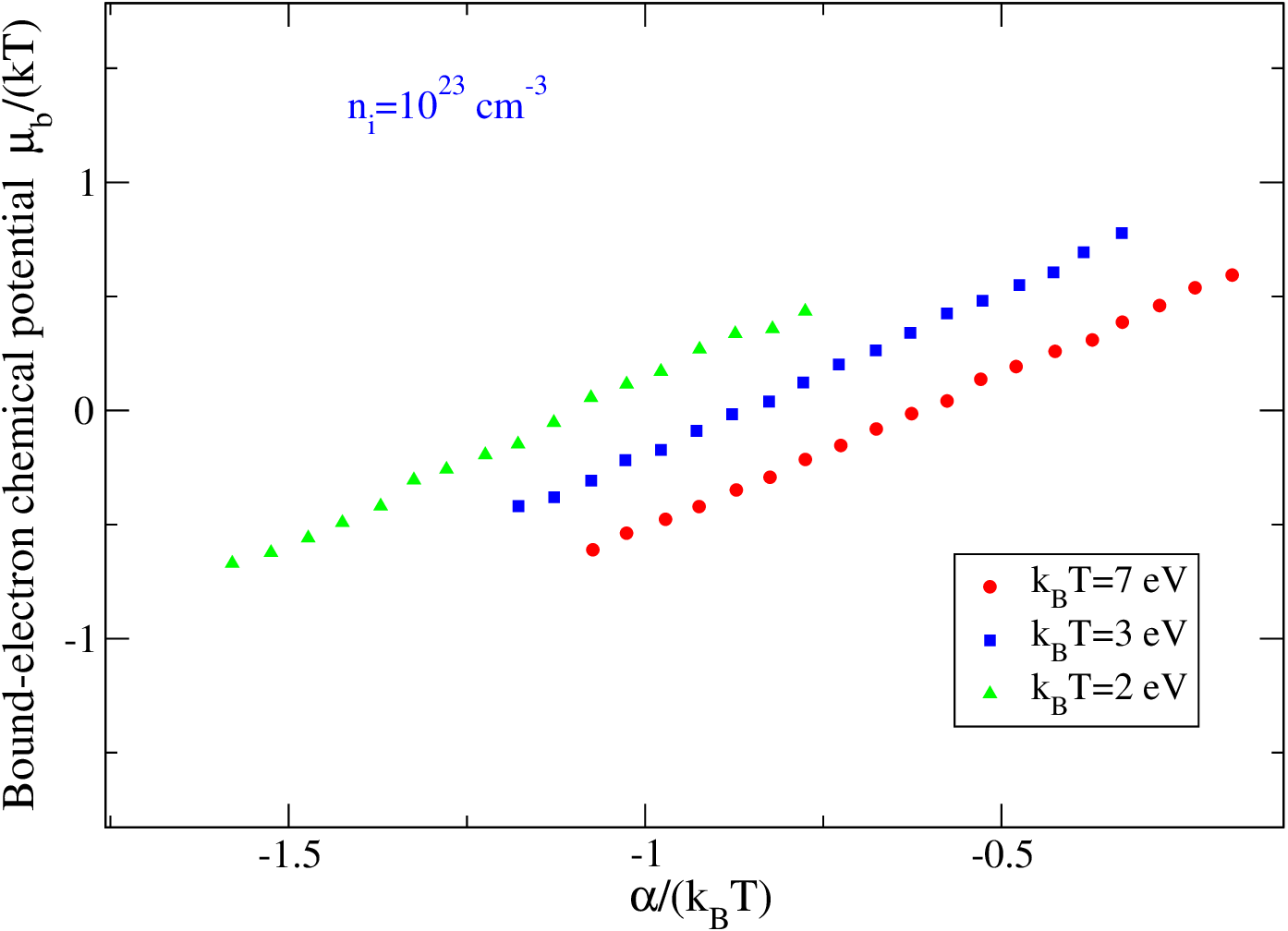} 
	\caption{Reduced bound-electron chemical potential $\mu_b/(k_BT)$ as a function of reduced parameter $\alpha/(k_BT)$ for $n_i$=10$^{23}$ cm$^{-3}$ and $k_BT$=2, 3 and 7 eV.}
	\label{13-14-15-16}
\end{figure}

Figures \ref{kalman_1019}, \ref{kalman_1021} and \ref{kalman_1023} represent the mean ionization of a hydrogen plasma for $n_i$=10$^{19}$, $n_i$=10$^{21}$ and $n_i$=10$^{23}$ cm$^{-3}$ respectively. We can see that, although the behaviour is similar (sharp increase before reaching a plateau), there are significant discrepancies between the results of Ref. \cite{Ying1989} and the Saha model, especially at low temperature. The differences increase with the density. Figure\,\ref{fig17} displays the reduced chemical potential $\mu_b/(k_BT)$ at $k_BT$=1 eV for $n_i$=10$^{21}$ and $n_i$=10$^{23}$ cm$^{-3}$ respectively, and figure \ref{13-14-15-16} for $n_i$=10$^{23}$ cm$^{-3}$ and $k_BT$=2, 3 and 7 eV, still in the case of a hydrogen plasma. The variation of the bound-electron chemical potential becomes more and more linear as the temperature increases. In addition, the origin of the difference between $\mu_b$ and $\alpha$ can been understood by analyzing the considering the ``effective'' bound-electron number
\begin{equation}
    \tilde{N}_b=\frac{32\pi^2}{3h^3}2m_ee^{3/2}\int_0^{r_0}drr^2\left[V(r)-V(r_0)\right]^{3/2}
\end{equation}
entering Eq. (\ref{c8}) and which is formally identical to the $N_b=Z-z$ at zero temperature. Since the definition of $N_b$ yields $\partial N_b/\partial z<0$, one can show that $\partial \tilde{N}_b/\partial z<0$ as well. As $z$ increases and $N_b$ decreases, the potential inside the ion departs from the highly screened atomic potential and tends to the pure Coulomb $Ze/r$ potential. This effect overcomes the simultaneous increase of $V(r_0)$ and both $V(r)-V(r_0)$ and $\tilde{N}_b$ increase. The difference between $N_b$ and $\tilde{N}_b$ resides in the fact that the Fermi distribution at finite temperature is quite different from the zero-temperature Heaviside distribution.

This idea of bound-electron chemical potential is also encountered, \emph{albeit} with a slightly different interpretation, in the superconfiguration formalism for the statistical computation of radiative opacity. In that framework, the bound-electron chemical potential plays the role of a Lagrange multiplier, associated to the preservation of the (integer) number of bound electrons in a super-configuration. The latter quantity is an ensemble of configurations, and is built by gathering some subshells inside supershells, such as
\begin{equation}
    S=(1s2s2p)^{10}(3s3p)^5(3d4s)^3(4p4d4f)^7
\end{equation}
where $(4p4d4f)^7$ means that one has to consider all the possibilities to make $(4p)^x(4d)^y(4d)^z$ with $x+y+z=7$ and of course $x\leq 6$, $y\leq 10$ and $z\leq 14$. This results in 35 possibilities, and finally $S$ contains 315 configurations. Such an approach was initially invented for opacity or emissivity calculations, but was also applied to equation of state \cite{Pain2002,Pain2012} or electrical resistivity calculations \cite{Pain2010}. The corresponding formalism enables one to investigate structural properties of plasmas beyond the average-atom model, which considers only an average configuration of the plasma (the subshells having non-integer populations). The average population of subshell $a$ (for instance $4d$) inside the supershell $\sigma=(4p4d4f)$ reads
\begin{equation}\label{pop3}
    \langle p_a\rangle=\frac{g_a}{1+\displaystyle\frac{\mathcal{U}_{Q_{\sigma}}\left(g^a\right)}{X_a\mathcal{U}_{Q_{\sigma}-1}\left(g^a\right)}}.
\end{equation}
where $Q_{\sigma}$ is the number of electrons in $\sigma$ and $\mathcal{U}_{Q_{\sigma}}(g)$ its partition function:
\begin{equation}
    \langle p_a\rangle=\frac{1}{\mathcal{U}_{Q_{\sigma}}(g)}\sum_{\substack{\{p_s\}\\\sum_{s=1}^Np_s=Q_{\sigma}}}g_a\prod_{s=1}^N\binom{g_s-\delta_{s,a}}{p_s-\delta_{s,a}}X_s^{p_s}.
\end{equation}
with $X_s=e^{\beta(\epsilon_s-\mu)}$, $\beta=1/(k_BT)$, $\epsilon_s$ is the energy of subshell $s$, $g_s$ its degeneracy, $\beta=1/(k_BT)$ and $\mu$ the chemical potential associated to the electro-neutrality of the plasma:
\begin{equation}
    N_{\mathrm{bound}}+N_{\mathrm{free}}=Z,
\end{equation}
$N_{\mathrm{bound}}$ being the number of bound electrons in the plasma and $N_{\mathrm{free}}$ the number of free electrons. The notation $g^a$ means that in the supershell $\sigma$, the degeneracy of the subshell $a$ was reduced by one ($g^a$ is sometimes called a ``reduced'' or ``shifted'' degeneracy). In some models, Eq. (\ref{pop3}) can be replaced by the Fermi-Dirac like form:
\begin{equation}
    \langle p_a\rangle=\frac{g_a}{1+\exp\left[\beta(\epsilon_a-\mu_{\sigma})\right]},
\end{equation}
where $\mu_{\sigma}$ is determined ensuring \cite{Pain2003}:
\begin{equation}
    \sum_{i\in\sigma}\langle p_i\rangle=Q_{\sigma}
\end{equation}
and is interpreted as a ``bound-electron chemical potential''. In a different context, it is worth mentioning that several physical ideas proposed by Ruoxian Ying and Kalman are reflected in many works on the average-atom models (see for instance \cite{Crowley1990}).

Finally, it is worth mentioning that, in their paper, Ruoxian Ying and Kalman also considered a Debye-type potential of the kind
\begin{equation}
    g_{ii}[r,V(r)]=\exp\left[-\frac{ZeV(r)}{k_BT}\right]-1.
\end{equation}

\section{On approximants of the Thomas-Fermi function $\Phi$}\label{sec4}

In the case where $e^{y-x}\ll 1$, the integral $J_n$ becomes
\begin{equation}
    J_n(x,x_0)\approx \int_x^{x_0}y^ndy=\frac{(x-x_0)^{n+1}}{n+1}
\end{equation}
and thus the equation (\ref{tfk}) turns into
\begin{equation}
    \psi''(x)=\frac{2a}{3}x\left[\frac{\psi(x)}{x}-\psi(1)\right]^{3/2}.
\end{equation}
If $\Psi(1)=0$, the latter equation boils down to 
\begin{equation}\label{ryktf}
    \psi''(x)=\frac{2a}{3}\frac{\psi(x)}{\sqrt{x}}.
\end{equation}

In the original Thomas-Fermi model \cite{Thomas1927,Fermi1927}, setting $V(r)-V_0=\Phi(r)Ze/r$, $x=r/r_0$ and $V_0=V(r_0)$ yields the famous Thomas-Fermi equation
\begin{equation}\label{TFEQ}
    \frac{d^2\Phi}{dx^2}=\frac{\Phi^{3/2}}{\sqrt{x}}
\end{equation}
with the boundary conditions $\Phi(0)=1$ and $\Phi(\infty)=0$. We can see that, rescaling the variable and redefining the function $\Psi$ correspondingly, Eq. (\ref{ryktf}) is of the same kind as Eq. (\ref{TFEQ}). 

\subsection{Solution of the Thomas-Fermi equation with boundary conditions $\Phi(0)=1$ and $\Phi(\infty)=0$}

From graphical considerations, Fermi deduced \cite{Ramnath1970}:
\begin{equation}
    \Phi(x)=1-Bx+\frac{4}{3}x^{3/2}
\end{equation}
with $B\approx 1.58$. In 1930, Baker obtained \cite{Baker1930}

\begin{equation}\label{bak}
    \Phi(x)=1-Bx+\frac{x^3}{3}-\frac{2}{15}Bx^4+\cdots +x^{3/2}\left[\frac{4}{3}-\frac{2}{5}Bx+\frac{3}{70}B^2x^2+\cdots\right]
\end{equation}
with $B\approx 1.588558$ \footnote{There is a typographical error in Ref. \cite{Epele1999}: $-2/5$ should be replaced by $2/5$, since Epele has chosen a convention different from Bakers's one \cite{Baker1930}, substituting $B$ by $-B$.}. It can be easily seen that Eq. (\ref{TFEQ}) has the particular solution
\begin{equation}
    \Phi(x)=\frac{144}{x^3},
\end{equation}
which vanishes asymptotically, as required by the second boundary condition. Clearly, it does not satisfy the condition at the origin, where this solution is singular. However, Sommerfeld \cite{Sommerfeld1932} achieved an analytical approximation to the exact solution of the Thomas-Fermi problem for neutral atoms by applying an asymptotic method that leads to the following result:
\begin{equation}
    \Phi(x)=\frac{144}{x^3}\left[1+\left(\frac{144}{x^3}\right)^{1/\lambda}\right]^{-\lambda} 
\end{equation}
with $\lambda=3.88$. Even though the Sommerfeld approximation has the right asymptotic behavior, it underestimates $\Phi(x)$ near the origin. The extended Sommerfeld approximation \cite{Sommerfeld1932} consists in writing
\begin{equation}
    \Phi(x)=\frac{1}{(1+z)^{\lambda_1/2}}
\end{equation}
where
\begin{equation}
    z=\left(\frac{x}{12^{2/3}}\right)^{\lambda_2}
\end{equation}
with $\lambda_1$ and $-\lambda_2$ solutions of
\begin{equation}
    \lambda^2-7\lambda-6=0
\end{equation}
and $\lambda_1=(7+\sqrt{73})/2\approx 7.7720018726587655839$ and $-\lambda_2=(7-\sqrt{73})/2=-\lambda_1/10$.
Although rational approximations (Pad\'e approximants) are probably the most accurate approximations \cite{Fernandez2011,Epele1999,Mavrin2021}, the variational approaches present often a good compromise between simplicity and accuracy. Roberts \cite{Roberts1968} used the trial function
\begin{equation}
    \Phi(x)=(1+\eta\sqrt{x})\,e^{-\eta\sqrt{x}},
\end{equation}
with $\eta=1.905$, followed a few years after by Csavinszky \cite{Csavinszky1973}
\begin{equation}
    \Phi(x)=(a_0\,e^{-\alpha_0 x}+b_0\,e^{-\beta_0 x})^2,
\end{equation}
the parameters being provided in Table \ref{tab1}, and Kesarwani and Varshni \cite{Kesarwani1981}
\begin{equation}
    \Phi(x)=(a\,e^{-\alpha x}+b\,e^{-\beta x}+c\,e^{-\gamma x})^2
\end{equation}
(see also Table \ref{tab1} for the parameters), while Wu used the following form \cite{Wu1982}:
\begin{equation}\label{wu}
    \Phi(x)=(1+m\sqrt{x}+nx)^2\,e^{-2m\sqrt{x}}
\end{equation}
with $m=1.14837$ and $n=4.0187$ 10$^{-6}$. A few years ago, Desaix \emph{et al.} \cite{Desaix2004} took advantage of the fact that the Euler-Lagrange equation corresponding to the Lagrangian
\begin{equation}
    L=\frac{1}{2}\left(\frac{d\Phi}{dx}\right)^2+\frac{2}{5}\frac{\Phi^{5/2}}{\sqrt{x}},
\end{equation}
i.e.,
\begin{equation}
    \frac{\partial L}{\partial\phi}-\frac{d}{dx}\frac{\partial L}{\partial\left(\displaystyle\frac{d\phi}{dx}\right)}=0
\end{equation}
is equivalent to the Thomas-Fermi equation. Inspired by the Sommerfeld form, the authors tried the function
\begin{equation}
    \Phi(x)=\frac{1}{\left[1+(kx)^{3/\alpha}\right]^{\alpha}},
\end{equation}
which parameters $k$ and $\alpha$ are obtained by solving
\begin{equation}
    \frac{d\langle L\rangle}{d\alpha}=0,
\end{equation}
where
\begin{equation}
    \langle L\rangle=\int_0^{\infty}Ldx.
\end{equation}
More recently, Oulne suggested the trial function \cite{Oulne2005,Oulne2011}\footnote{The is a typographical error in Oulne's paper concerning the Wu function, which should be as Eq. (\ref{wu}).}:
\begin{equation}
    \frac{1}{\left[1+(kx)^{3/\beta}\right]^{\alpha}},
\end{equation}
with $k=0.4835$, $\alpha=2.098$ and $\beta=0.9238$, which is a generalization of Desaix's work. For the Thomas–Fermi model of a multi-electron atom and a positively charged ion, highly efficient computational algorithms were constructed by Pikulin that solved the problem for an atom (that is, the boundary value problem on the half-line $0<x<\infty$) and found the derivative of this solution with any prescribed accuracy at an arbitrary point of the half-line \cite{Pikulin2019}. The results are based on an analytic property of a special Abel equation of the second kind to which the original Emden–Fowler equation 
\begin{equation}\label{ef}
    \left(x^py'\right)'\pm x^{\sigma}y^n=0
\end{equation}
reduces. More precisely, the solutions that pass a modified Painlev\'e test at a nodal singular point of the equation can be represented by a convergent power series. The Thomas-Fermi equation corresponds to the minus-sign case in $\pm$, and $p=0$, $\sigma=-1/2$ and $n=3/2$ in Eq. (\ref{ef}). 

\begin{figure}[!ht]
  \centering
  \includegraphics[width=8cm,angle=\anglefig,scale=\scalefig]{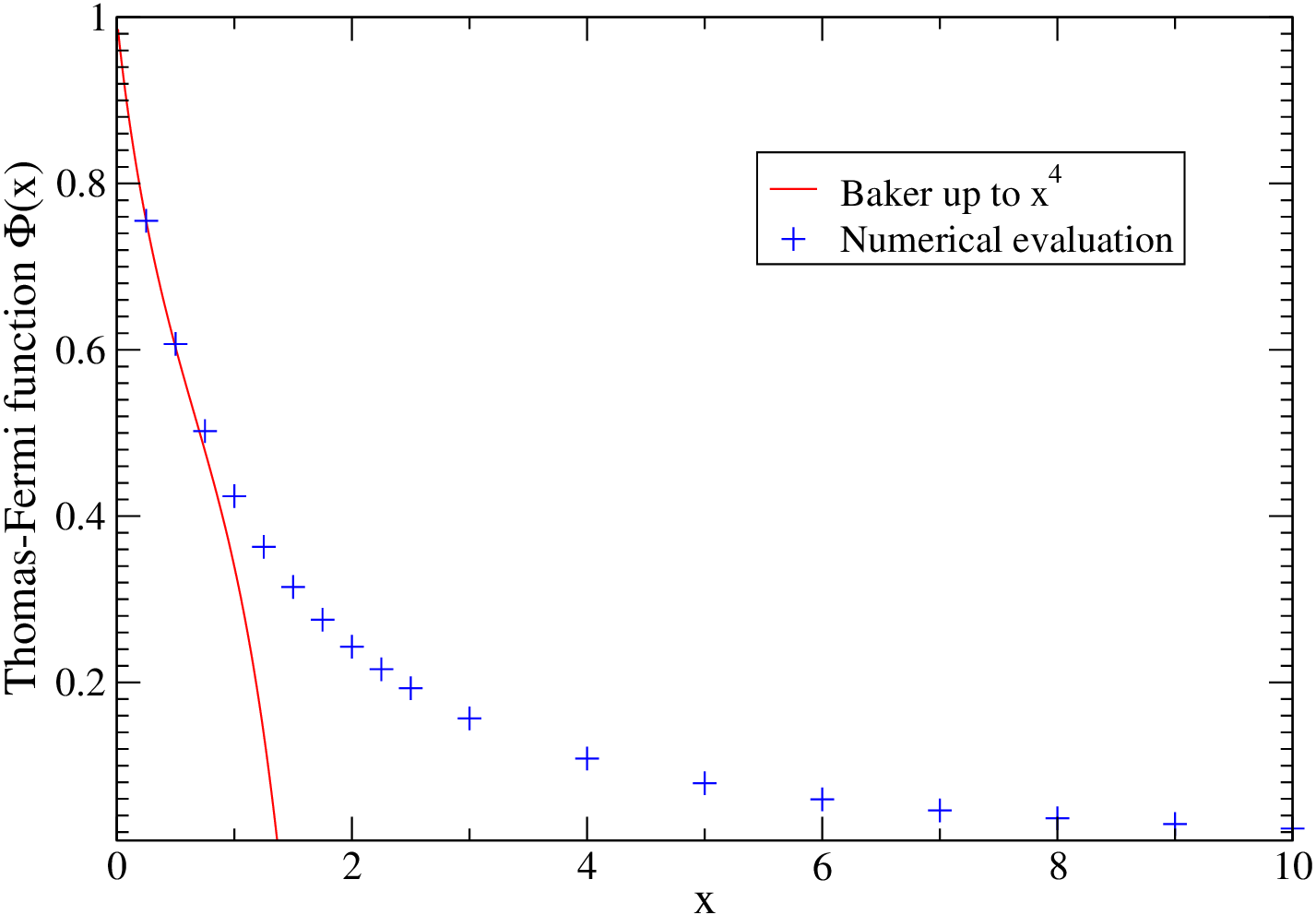}
  \caption{(Color online) Approximate formulas of Baker up to $x^4$ (see Eq. (\ref{bak})) \cite{Baker1930} of the Thomas-Fermi function $\Phi$ compared to high-precision numerical evaluation \cite{Parand2017}. \label{baker_lin}}
\end{figure}

\begin{figure}[!ht]
  \centering
  \includegraphics[width=8cm,angle=\anglefig,scale=\scalefig]{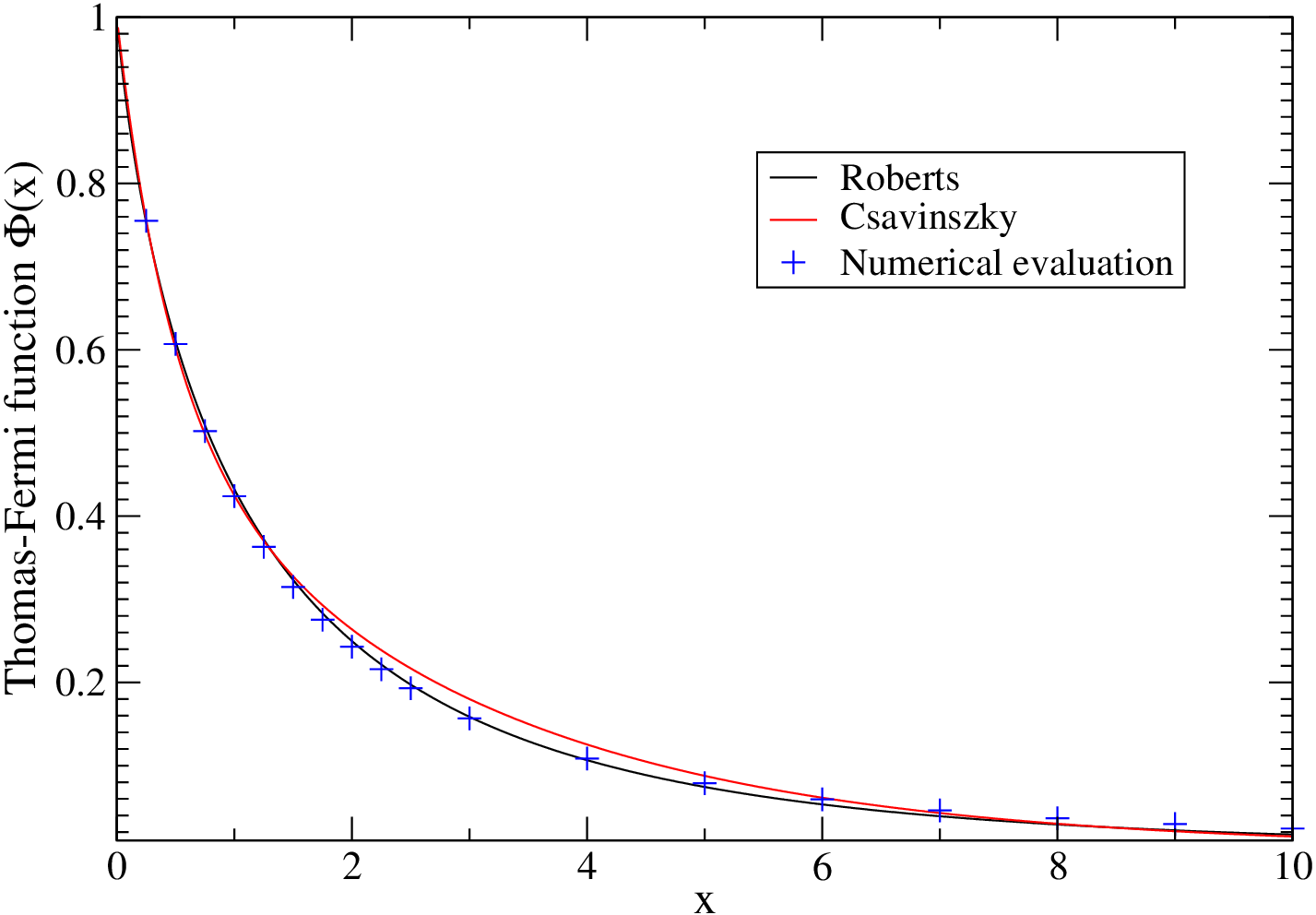}
  \caption{(Color online) Approximate formulas of Roberts \cite{Roberts1968} and Csavinszky \cite{Csavinszky1973} of the Thomas-Fermi function $\Phi$ compared to high-precision numerical evaluation \cite{Parand2017}. \label{first_lin}}
\end{figure}

\begin{figure}[!ht]
  \centering
  \includegraphics[width=8cm,angle=\anglefig,scale=\scalefig]{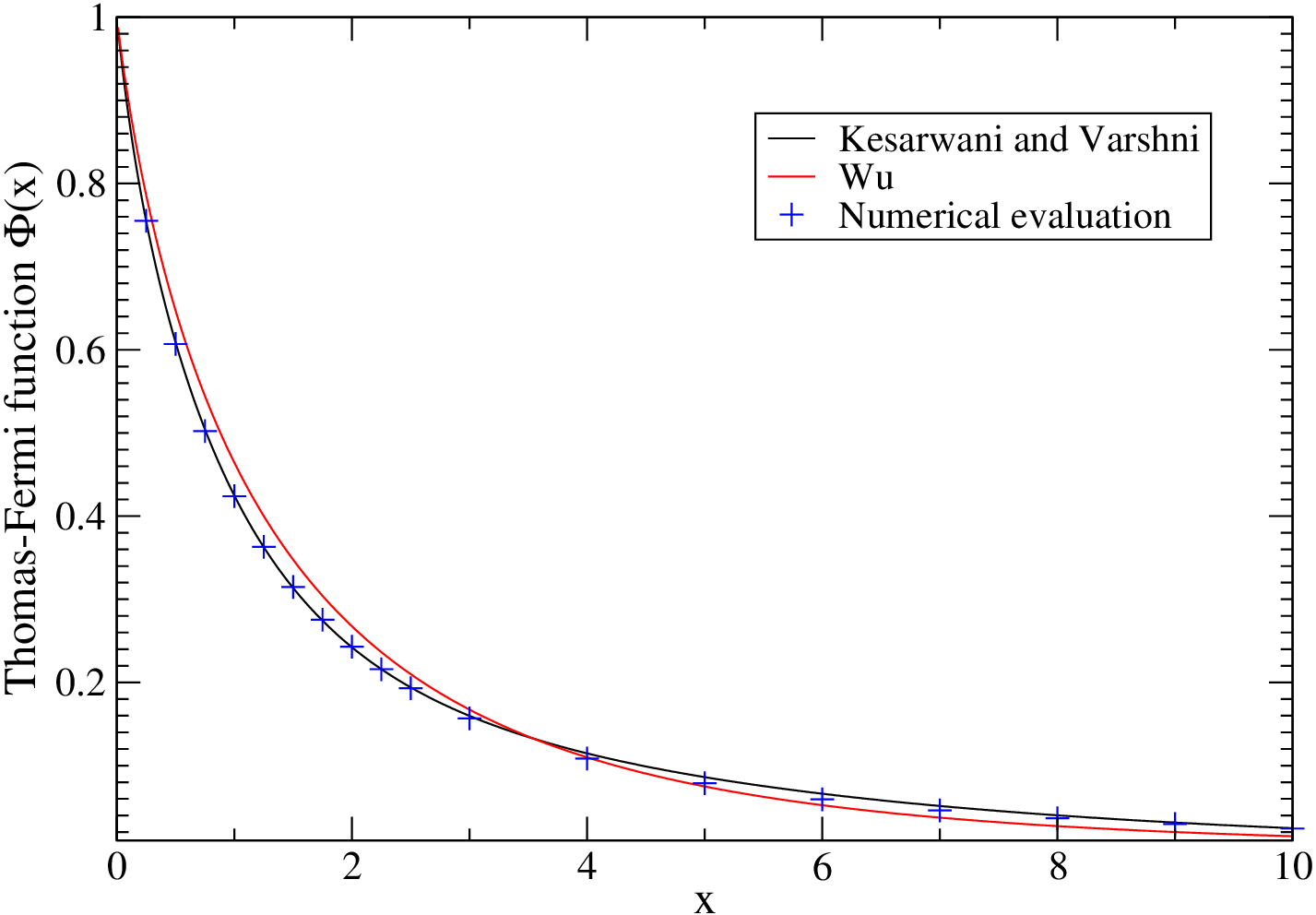}
  \caption{(Color online) Approximate formulas of Kesarwani and Varshni \cite{Kesarwani1981} and Wu \cite{Wu1982} of the Thomas-Fermi function $\Phi$ compared to high-precision numerical evaluation \cite{Parand2017}. \label{second_lin}}
\end{figure}

\begin{figure}[!ht]
  \centering
  \includegraphics[width=8cm,angle=\anglefig,scale=\scalefig]{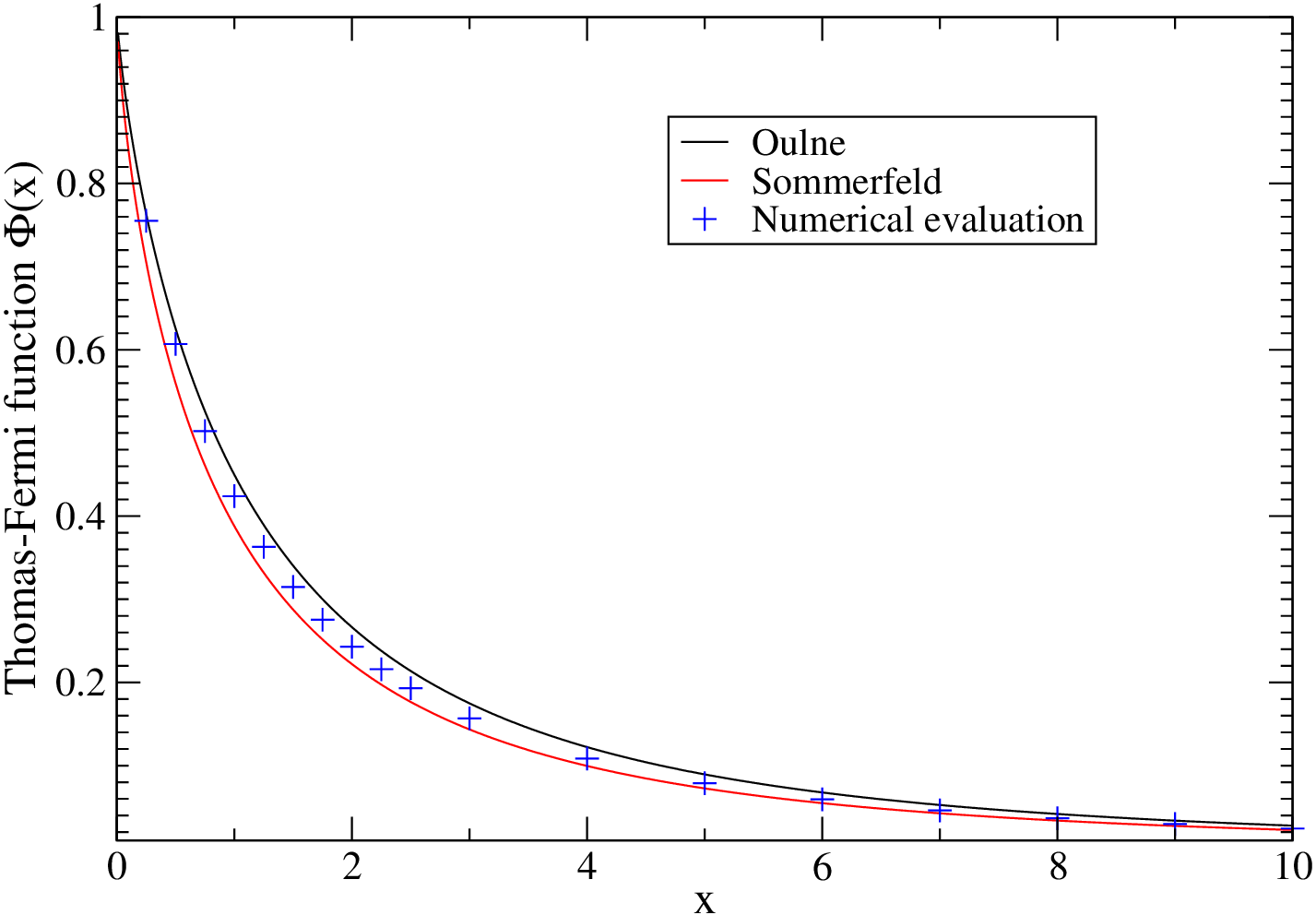}
  \caption{(Color online) Approximate formulas of Oulne \cite{Oulne2005,Oulne2011} and Sommerfeld \cite{Sommerfeld1932} of the Thomas-Fermi function $\Phi$ compared to high-precision numerical evaluation \cite{Parand2017}. \label{third_lin}}
\end{figure}

As shown in figure \ref{baker_lin}, the Baker expansion up to $x^4$ is valid only for very small values of $x$. Beyond that, the expansion would have to be performed up to very high powers in order to cover the range considered here (i.e., $x\in[0,10]$) and problems due to the summation of alternating-sign terms would occur.

Figure \ref{first_lin} represents the approximants of Roberts \cite{Roberts1968} and Csavinszky \cite{Csavinszky1973} compared to recent highly accurate numerical values \cite{Parand2017}. We can see that Roberts' formula has a high accuracy up to $x=4$, but tends to overestimate $\Phi(x)$ for higher values. Csavinszky works fine up to $x=2$, then overestimates $\Phi(x)$ between 2 and 7, and is too low afterwards. Figure \ref{second_lin} represents the approximants of Kesarwani and Varshni \cite{Kesarwani1981} and Wu \cite{Wu1982} compared to the above mentioned accurate numerical values \cite{Parand2017}. The former has a high accuracy, except around $x=5$ (its values are a bit too high), and reproduces well the limiting cases (close to $x=0$ and $x=10$). The Wu representation is slightly too high up to 4, and much too low afterwards. As can be seen in figure \ref{third_lin}, the Sommerfeld approximation \cite{Sommerfeld1932} is always a bit too low, but works globally well and the Oulne formula \cite{Oulne2005,Oulne2011} is accurate for small $x$ but significantly too high otherwise.

It is worth emphasizing that the formulas considered in the present work may not be the best expressions developed by the authors. We have chosen a few formulas for their simplicity and quality, but this study is neither exhaustive nor representative of all the efforts made by the authors to obtain a good representation of $\Phi(x)$.

\begin{table}
    \begin{center}
        \begin{tabular}{l|l}\hline
Csavinszky & Kesarwani and Varshni \\\hline\hline
$a_0$=0.7218337 & $a$=0.52495\\
$\alpha_0$=0.0.1782559 & $\alpha$=0.12062\\
$b_0$=0.2781663 & $b$=0.43505\\
$\beta_0$=1.759339 & $\beta$=0.84795\\
  & $c$=0.04\\
  & $\gamma$=6.7469\\\hline
        \end{tabular}
        \caption{Parameters entering the approximants of Csavinszky \cite{Csavinszky1973} Kesarwani and Varshni \cite{Kesarwani1981} for the function $\Phi(x)$.}\label{tab1}
    \end{center}
\end{table}

Zhu \emph{et al.} published an adaptive algorithm to solve the problem by means of the moving mesh finite-element method \cite{Zhu2012}. Jovanovic \emph{et al.} solved Thomas–Fermi equation by applying a spectral method using an exponential basis set in a semi-infinite domain \cite{Jovanovic2014}. Fatoorehchi and Abolghasemi have employed a newly analytical scheme that relies on the improved differential transform method and the Pad\'e-approximant technique to obtain an explicit series solution to the Thomas–Fermi equation \cite{Fatoorehchi2014}. Parand and Delkhosh derived accurate solution of the Thomas–Fermi equation using the fractional order of rational Chebyshev functions \cite{Parand2017}. Their paper contains also an extensive review of all the theoretical works related to the Thomas-Fermi equation (approximants, numerical solving, derivative at the origin, \emph{etc.}).

\subsection{Ion in plasma ($\Phi(x_0)=0$): improvements and effect of mean ionization}

In the case of an ion in plasma, the boundary conditions are now $\Phi(0)=1$ and $\Phi(x_0)=0$ (instead of $\Phi(\infty)=0$). Even though the Sommerfeld approximation has the right asymptotic behavior, it underestimates $\Phi(x)$ near the origin. The extended Sommerfeld approximation \cite{Sommerfeld1932} consists in writing
\begin{equation}\label{som}
    \Phi(x)=\frac{1}{(1+z)^{\lambda_1/2}}\left[1-\left(\frac{1+z}{1+z_0}\right)^{\lambda_1/\lambda_2}\right]
\end{equation}
where $z$, $\lambda_1$ and $-\lambda_2$ are the same as in the previous section and
\begin{equation}
    \frac{z_0}{(1+z_0)^{\lambda_1/2+1}}=\frac{q}{\lambda_1},
\end{equation}
where $q=(Z-N)/Z$, $N$ being the number of bound electrons\footnote{As stated by Mavrin and Demura, the expression in the square bracket of the right-hand side of Eq. (\ref{som}) can be improved if the exact numerical values are used for the function $\Phi_0$ \cite{Gombas1949}: $z_0\Phi_0(x_0)=q(1+z_0)/\lambda_1$ \cite{Mavrin2021}.}. Kobayashi proposed a series expansion \cite{Kobayashi1957}. He also suggested to use the Fermi expression \cite{Fermi1931,Kobayashi1956}:
\begin{equation}
    \Phi(x)=\Phi_0(x)+k(q)\eta_0(x).
\end{equation}
with
\begin{equation}
    k(q)=-\frac{\Phi_0(x_0)}{\eta_0(x_0)}
\end{equation}
as well as
\begin{equation}
    \eta_0(x)=\left[\Phi_0(x)+\frac{x}{3}\Phi_0'(x)\right]\int_0^x\frac{du}{\left[\Phi_0(u)+\displaystyle\frac{u}{3}\Phi_0'(u)\right]^2}
\end{equation}
and $k=-0.083\,q^3$ \cite{Fermi1934} combined with expressions of $\eta_0$ and $\eta_0'$ by Gombas \cite{Gombas1949}, Miranda \cite{Miranda1934} and Fermi and Amaldi \cite{Fermi1934}. A few years later, Kobayashi refined his estimate and prescribed $k=-0.0542\,q^{2.83}$ \cite{Kobayashi1959}. Note that Majorana expressed the solution of the Thomas-Fermi equation in terms of one quadrature only \cite{Esposito2002}. Mavrin and Demura recently published an interesting approach \cite{Mavrin2021}, using the Mason formulas \cite{Mason1964}:
\begin{equation}
    \Phi_0(x)=\left(\frac{1+1.81061\,x^{1/2}+0.60112\,x}{1+1.81061\,x^{1/2}+1.39515\,x+0.77112\,x^{3/2}+0.21465\,x^2+0.04793\,x^{5/2}}\right)^2
\end{equation}
as well as
\begin{equation}
    \eta_0(x)=\exp\left(z+0.3837\,z^2+0.0892\,z^3-0.0170\,z^4\right)-1
\end{equation}
with $z=\ln(1+x)$. Mavrin and Demura suggested, for a given ionization $q$, to approximate $x_0=12^{2/3}\,z_0^{1/\lambda_2}$ (solution of $-x_0\Phi'(x_0)=q$) as \cite{Mavrin2021}:
\begin{equation}
    \frac{10.232}{q^{1/3}}\left(1-0.917\,q^{0.257}\right)\qquad\mathrm{when}\qquad q\leq 0.45
\end{equation}
and
\begin{equation}
    2.960\left(\frac{1-q}{q}\right)^{2/3}\qquad\mathrm{when}\qquad q > 0.45.
\end{equation}

\section{Conclusions}\label{sec7}

Personally, I have fond memories of Gabor Kalman. Although I know him mainly through his work and have never been a close friend, I had several opportunities to discuss physics and much more general topics with him at ``Strongly Coupled Coulomb Systems'' and ``Physics of Non-ideal Plasmas'' conferences. He was very important in the community, and made major contributions to plasma physics. Even if the subject of this article may seem anecdotal in the light of his scientific output, and even if it did not encounter as much resonance in the community as the work with Ken Golden on quasi-particles, response functions and plasmon dispersion \cite{Kalman1990,Golden2000}, it deserves, in my humble opinion, to be brought to the fore. The work of Gabor and his colleagues led to major advances in the understanding and modeling of correlations and the dispersion of plasma oscillations affecting the average and fluctuating fields in a strongly coupled plasma,  as well as in the establishment of methods for the calculation of the degree of ionization and of the shift of energy levels of an ion embedded in a dense plasma. In the last part of the article, we proposed a (non-exhaustive) state of the art of analytical representations of the Thomas-Fermi function. Such approximants can be generalized to the function $\Psi$ entering the above mentioned formalism of Ruoxian Ying and Kalman, related to the finite-temperature Thomas-Fermi model.

\section*{Acknowledgments}
I am indebted to Marlene Rosenberg, Jim Dufty, Zolt\'an Donk\'o, and Peter Hartmann for organizing the special issue dedicated to the memory of Gabor J. Kalman and Kenneth I. Golden.

\end{document}